# e-person Architecture and Framework for Human-AI Co-adventure Relationship

Kanako Esaki[1][0000-0002-3269-9130], Tadayuki Matsumura[1][0009-0009-1907-1996],
Yang Shao[1][0009-0009-0655-9562] and Hiroyuki Mizuno[1][0000-0002-1213-9021]

[1] Hitachi Ltd., Tokyo, Japan

`kanako.esaki.oa@hitachi.com`

**Abstract.** This paper proposes the e-person architecture for constructing a unified and incremental development of AI ethics. The e-person architecture takes the reduction of uncertainty through collaborative cognition and action with others as a unified basis for ethics. By classifying and defining uncertainty along two axes - (1) first, second, and third person perspectives, and (2) the difficulty of inference based on the depth of information - we support the development of unified and incremental development of AI ethics. In addition, we propose the e-person framework based on the free energy principle, which considers the reduction of uncertainty as a unifying principle of brain function, with the aim of implementing the e-person architecture, and we show our previous works and future challenges based on the proposed framework.

## 1 Introduction

With the rapid spread of AI into everyday life, discussions on AI ethics have become more active [1-7]. Assuming that the application of AI in everyday life will continue to expand, it is expected that AI ethics AI will become more important as people strive to lead enriched lives while using AI. So far, there have been many discussions on ethical concerns related to AI, focusing on philosophy and sociology, such as the singularity and the threat of AI [8,9]. On the other hand, the current discussion is characterized by the fact that with the practical application and rapid proliferation of technologies such as self-driving cars and chatbots, more practical discussions have become active, such as those on regulation [10] and their technical implementation [11-14]. In other words, the ethical issues of AI are not hypothetical problems in the distant future, but urgent and actual problems that need to be addressed, and they are unavoidable not only for philosophers and sociologists, but also for AI service providers and AI engineers.

Current AI ethics practice is active in the approach of decomposing ethical issues into specific multiple problems and solving each problem individually. For example, in [15], the five important perspectives for developing responsible AI are explained as explainability, fairness, robustness, transparency, and privacy, and the direction of practical efforts is shown. In particular, the discussion and development of AI ethics has been stimulated by breaking them down into quantitative problems. In engineering, where the basic approach is to improve and optimize quantitative indicators, it is a useful approach to define problems with vague quantitative indicators, such as ethical issues, as multiple problems that can be quantified, and to develop solutions for each of them. For example, various quantitative evaluation indicators have been proposed for fairness, and methods for improving them are actively being developed [16-17].

On the other hand, there do not seem to be many opportunities to discuss and develop ethical issues from an engineering perspective in a unified way. For example, there has not been sufficient progress in deepening the engineering-based unified understanding of the question "What is ethical AI?". Issues such as fairness and safety are obviously important in the practice of ethical AI. However, we believe that the difficulty of AI ethics issues lies in the conflict between multiple issues. For example, the Trolley Problem is a well-known hypothetical problem that demonstrates the difficulty of making ethical decisions, however in the case of automated driving, the Trolley Problem is not a hypothetical problem, but rather a real-world conflict between the safety of the driver and the fairness of human life. For this reason, we believe it is important to develop and implement solutions based on a unified view of what AI ethics is, rather than addressing ethical issues from multiple individual perspectives.

In addition, when tackling big issues such as ethics, it is important in the practice of development not only to create a unified picture, but also to divide it into appropriate stages and levels, and to accumulate results step by step toward the final goal. The same problem can be seen in the fields of automated driving and artificial general intelligence (AGI). In the development of automated driving, the meaning of "automated" is broad, so it is important to have a common understanding of "automated" in order to have effective discussions among practitioners. In the field of automated driving, the Society of Automotive Engineers (SAE) has proposed a system of levels of automated driving[18], which enables developers to have more effective discussions by clarifying what level of automated driving they are discussing, and also enables the incremental improvement of automated driving. Similarly, in the development of AGI, DeepMind has proposed levels of AGI[19] to enable more effective discussions between AGI researchers and practitioners, as the meaning of "general" is broad and ambiguous.

We believe that the importance of the AGI and Autonomous Driving levels lies in the fact that they do not break down grand goals into multiple goals, but rather capture the problem from a unified perspective and define the stages of development of the problem from that unified perspective. For example, at the AGI level, the grand goal of "general" is captured from the two perspectives of "autonomy" and "performance," and the development stages are defined. This paper proposes a new practical approach to AI ethics problems that (1) draws a picture of AI ethics from a unified perspective, rather than breaking down ethical issues into multiple individual problems, and (2) breaks down (and clearly specifies) this picture into stages (levels) from a unified perspective. Specifically, we propose an e-person architecture that captures AI ethical issues from a unified perspective and defines their levels. The e-person architecture reinterprets the e-person concept proposed by philosopher Deguchi [20,21] from an engineering perspective into a form that can be put into practice. The e-person architecture sees the basis of ethics as trust in people, and classifies and organizes this as a process of reducing uncertainty that can be manipulated from an engineering viewpoint. In particular, it is characterized by its method of trying to reduce uncertainty not through AI alone, but through interaction and communication with humans and society. In contrast to the conventional many discussions of AI ethics to date have focused on the ethics of AI (Ethics "of" AI) and the impact of AI ethics on human ethics (Ethics "by" AI), the e-person architecture takes a different approach, viewing ethics as something that should be created jointly by humans and AI (Ethics "with" AI) and focusing on the bidirectional relationship between humans and AI. In other words, it is characterized by reconceptualizing the relationship between humans and AI from the conventional unidirectional relationship of humans using AI as a tool to a bidirectional relationship of humans and AI as co-adventurers.

In Section 2, this paper organizes existing discussions and issues regarding existing AI ethics. Next, in Section 3, we propose the e-person architecture with the aim of solving the problems in the existing discussions. Finally, in Section 4, we propose a framework for implementing the e-person architecture in concrete terms, and we introduce our past efforts and future issues in line with the proposed framework.

## 2    Related Research

### 2.1    Philosophical Exploration of Artificial Intelligence: From Early Speculations to Modern Ethical Challenges

Artificial intelligence (AI) has been a subject of discussion primarily in the field of philosophy since its inception at the Dartmouth Conference in 1956. Early philosophical discussions mainly focused on the definition of AI, its possibilities, and its relationship with human intelligence. As AI technology rapidly developed, philosophers gradually turned their attention to ethical issues surrounding AI, its moral status, controllability, and social impact.

This section organizes the most influential philosophical literature on AI from the past 20 years and analyzes its development in chronological order.

**The Moral Status of AI: The Starting Point of AI Ethics.** The ethical exploration of artificial intelligence can be traced back to Floridi and Sanders (2004) [22], who published "On the Morality of Artificial Agents." This paper was the first to systematically analyze the moral status of artificial agents. From the perspective of information ethics, they argued that while AI is not a moral agent in the traditional sense, it should be considered a moral patient deserving ethical consideration in certain contexts. This theory became the foundation for subsequent discussions on AI ethics and responsibility. Two years later, Moor (2006) [23], in the paper "The Dartmouth College Artificial Intelligence Conference: The Next Fifty Years," reflected on the development of AI and outlined philosophical challenges for the next half-century. He raised the issue of "AI's moral agency," posing the question of how ethical responsibility should be defined if AI can make autonomous decisions. This issue remains a central topic in AI governance and legal frameworks today.

**The Rise of Superintelligence and the AI Control Problem.** Entering the 2010s, as AI capabilities significantly improved, philosophers began to focus on the existential risks and control challenges posed by AI. Bostrom (2014) [24], in the book Superintelligence: Paths, Dangers, Strategies, systematically analyzed the risks posed by superintelligence. He introduced the "control problem" and the "value alignment problem," questioning whether AI, if it surpasses human intelligence, can be controlled and made to act in accordance with human values. These issues have become core challenges in the field of AI safety, prompting institutions like OpenAI and DeepMind to intensify research on AI governance. After Bostrom raised concerns about the threats of superintelligence, Russell (2019) [25], in Human Compatible: Artificial Intelligence and the Problem of Control, further developed the concept of human-compatible AI. He argued that future AI design should not merely pursue optimal decision-making but should center on human preferences and values. This idea has driven the development of human-centered AI and influenced policymakers' approaches to AI regulation.

**Deep Integration of Philosophy and AI Technology.** As AI technology entered the era of deep learning, discussions between philosophers and computer scientists became increasingly intertwined. At the same time, AI ethics research entered a data-driven phase. Cappelen and Dever (2021) [26], in Making AI Intelligible: Philosophical Foundations, argued that AI explainability is central to human-machine coexistence. They pointed out that humans should not blindly trust black-box models but must understand the logic behind AI decisions. This perspective has influenced the field of explainable AI, promoting the evolution of AI models toward greater transparency. Gao et al. (2024) [27], in "AI Ethics: A Bibliometric Analysis, Critical Issues, and Key Gaps," used large-scale data analysis to uncover the main trends in AI ethics research over the past 20 years. They found that AI ethics research has shifted from early discussions of moral agency to specific issues such as fairness, transparency, and algorithmic bias. This reflects the growing integration of AI technology into social structures. Finally, Velthoven and Marcus (2024) [28], in "Problems in AI, Their Roots in Philosophy, and Implications for Science and Society," pointed out errors in knowledge growth within AI philosophy. Citing the theories of Karl Popper and David Deutsch, they argued that current AI "learning" is essentially pattern recognition, not true knowledge creation. This view challenges mainstream AI cognitive models and provides a new philosophical framework for future AI development.

From the 2004 discussions on the moral status of AI to the 2024 deep reflections on explainability and knowledge growth, philosophical research on artificial intelligence has evolved from theoretical construction to technological integration. Early debates primarily focused on whether AI could be granted moral status, while the 2010s centered on the risks of superintelligence and human-machine relationships. Entering the 2020s, AI philosophical research became more practical, revolving around ethics, fairness, and explainability, driving progress in policymaking and technological optimization. In the future, with the development of generative AI, neuro-symbolic AI, and self-evolving AI, philosophy is expected to continue playing a crucial role in AI governance, value alignment, and models of human-machine coexistence.

## 2.2 The Intersection of Artificial Intelligence and Sociology: A Review of Recent Research and Development Trends

The development of artificial intelligence (AI) not only drives technological progress but also profoundly impacts social structures, economic systems, and human-machine relationships. In recent years, as AI capabilities have grown increasingly powerful, the field of sociology has engaged in a series of significant studies on AI ethics, social adaptability, cooperation mechanisms, and economic effects. This section organizes key studies from the past decade on the coexistence of AI and human society and analyzes their development in chronological order.

**Early Explorations: Frameworks for AI's Social Impact and Cooperation Mechanisms.** The social impact of AI was initially explored primarily around ethics, cooperation mechanisms, and its role in socioeconomics. Benjamins and Salazar (2020) [29], in "Towards a Framework for Understanding Societal and Ethical Implications of Artificial Intelligence," constructed a framework for socio-ethical analysis, exploring core issues such as AI fairness, transparency, and accountability, laying the groundwork for subsequent AI governance research.

**Deepening Research: AI's Impact on Labor Markets, Social Inequality, and Policy.** With the advancement of AI technology, sociologists began to focus on its profound socioeconomic effects. Deranty and Corbin (2024) [30], in "Artificial Intelligence and Work: A Critical Review of Recent Research from the Social Sciences," systematically reviewed research on the relationship between AI and labor within the social sciences, analyzing how automation transforms labor market structures and noting that while AI may reduce jobs, it also has the potential to create new forms of work. Capraro et al. (2024) [31], in "The Impact of Generative Artificial Intelligence on Socioeconomic Inequalities and Policy Making," further explored the effects of generative AI on socioeconomic inequality. They pointed out that the proliferation of AI could exacerbate disparities in wealth and resource distribution, while simultaneously posing higher governance demands on policymakers.

**Co-existence and Adaptation: AI's Dynamic Role in Social Systems.** Recent research has focused on the long-term symbiotic relationship between AI and human society. Pedreschi et al. (2024) [32], in "Human-AI Coevolution," proposed the theory of "human-AI coevolution," arguing that AI is not merely a technological tool but an intelligent system capable of forming dynamic cooperative relationships with human society. This theory offers a new perspective on human-machine hybrid intelligence, asserting that while AI must adapt to human society, human society must also adapt to AI's influence. Meanwhile, Mahmud et al. (2023) [33], in "A Study of Human-AI Symbiosis for Creative Work: Recent Developments and Future Directions in Deep Learning," focused on how AI is applied to creative work, analyzing how AI collaborates with humans in fields like art and writing. They argued that AI's role in creative industries should be seen not as a replacement but as an augmentative collaborative partner. Recent studies have further deepened the role of AI in the social sciences. Xu et al. (2024) [34], in "AI for Social Science and Social Science of AI: A Survey," explored how AI functions both as a tool for social science research and as a subject of study. They argued that AI not only drives methodological innovation in social research but also has the potential to influence social structures and individual behavior. Solís et al. (2023) [35], in "The Impact of Artificial Intelligence on Higher Education: A Sociological Perspective," analyzed AI's impact on higher education, exploring its potential effects on automated assessment, personalized learning, and educational support, reflecting AI's profound role in education and knowledge dissemination.

Sociological research on AI has evolved from "human-machine cooperation and ethics" to "impacts on economic policy" and finally to "symbiosis and adaptation." Currently, AI's role has shifted from being a tool to becoming part of an intelligent ecosystem, bringing methodological transformations to social science research itself. In the future, research is expected to advance on how AI shapes social governance models, influences cultural values, and deepens the theoretical framework of human-machine hybrid intelligence.

### 2.3   Engineering Research on Artificial Intelligence: The Evolution of Safe Coexistence and Human-AI Collaborative Design in the AGI Era

In response to the emerging reality of AGI (Artificial General Intelligence) and its accompanying safety challenges, engineering research on human-AI collaboration has transitioned from early unidirectional technical support to bidirectional, complementary symbiosis. This section organizes key literature along a timeline, dividing it into two phases: early exploration of basic interactions and safety mechanisms and the shift toward bidirectional collaboration and safe coexistence in the AGI era.

**Exploration of Basic Interactions and Safety Mechanisms.** Breazeal [36], in the 2003 work Toward Sociable Robots, first proposed the concept of social robots capable of human-like emotional expression and natural interaction, laying the emotional foundation for human-AI interaction. During the same period, Dzindolet et al. [37], in "The Role of Trust in Automation Reliance," experimentally verified the role of transparency in automation systems for building user trust, revealing the moderating effect of trust in automated decision-making. In 2004, Lee and See [38], in "Trust in Automation: Designing for Appropriate Reliance," systematized a methodology for designing trust in human-AI collaboration to prevent both overreliance and excessive skepticism. In 2008, Goodrich and Schultz [39], in "Human-Robot Interaction: A Survey," provided a comprehensive review of theories and practices for achieving safe and efficient coexistence in industrial and service robots. Gray and Wegner [40], in their 2012 paper "Feeling Robots and Human Zombies: Mind Perception and the Uncanny Valley," analyzed human psychological responses to attributing minds to robots and the "uncanny valley" phenomenon, serving as a crucial warning for human-centered design. In 2016, Amodei et al. [41], in "Concrete Problems in AI Safety," discussed specific challenges in AI safety from an engineering perspective, promoting the development of safety strategies for human-AI coexistence in high-risk environments.

**The Shift Toward Bidirectional Collaboration and Safe Coexistence in the AGI Era.** In 2019, Amershi et al. [42], in "Guidelines for Human-AI Interaction," systematized human-centered interaction design guidelines based on large-scale user studies, marking a fundamental shift from unidirectional technical support to bidirectional complementary collaboration. In 2020, Prakash and Mathewson [43], in "Conceptualization and Framework of Hybrid Intelligence Systems," constructed a theoretical framework for hybrid intelligence systems that integrate human experts and AI to address real-world challenges. In 2023, Fabri et al. [44], in "Disentangling Human-AI Hybrids – Conceptualizing the Interworking of Humans and AI-Enabled Systems," systematized the components and interaction mechanisms of human-AI hybrid systems from a hybrid intelligence perspective, providing a theoretical foundation for building efficient bidirectional collaboration systems. In 2024, Holter and El-Assady [45], in "Deconstructing Human-AI Collaboration: Agency, Interaction, and Adaptation," proposed a new collaboration framework based on three dimensions—agency, interaction, and adaptability—reflecting the increasing sophistication and precision of research on safe coexistence and collaborative design in the AGI era.

## 2.4 Summary

Up to this point, this section has provided an overview of related research in the fields of philosophy, sociology, and engineering concerning the ethical concerns of AI. To summarize the situation in each field: In philosophy, discussions have evolved from conceptual and speculative issues, such as moral agency and the singularity, to more concrete and practical concerns in recent years, such as responsibility in the event of accidents and AI governance, driven by progress in AI deployment. Similar to recent discussions in philosophy, in sociology, there are active discussions on specific issues like AI's impact on labor markets and economic inequality. A key feature of sociological discussions is that AI is not viewed as existing outside of human society but as one of its constituent elements, treating AI-related problems as intrinsic to society rather than external challenges. Likewise, in engineering, AI is also seen as an intrinsic societal issue, with active discussions on addressing it as a bidirectional problem based on human-AI interaction.

From these trends in related research, it is believed that a fundamental approach to AI ethics involves viewing the issues as intrinsic to society and focusing on bidirectional human-AI interactions for problem-solving. In other words, in the era of AI deployment, it is considered crucial to propose new AI models aimed at building a new relationship between humans and AI, envisioning a symbiotic society. These new AI models are expected to face the challenge of flexible and dynamic adaptation to diverse societies and cultures. Furthermore, realizing a new relationship between humans and AI through these models is thought to lead not only to simple technological advancements in AI but also to the creation of important research fields that innovate the knowledge generation process in society.

## 3      Proposal of e-person Architecture

### 3.1      Approach for Building e-person Architecture

This paper proposes an alternative AI architecture to the conventional unidirectional use of AI as a tool by humans, in which artificial agents live and form society together with us as equal partner: the e-person architecture. The e-person architecture reinterprets the e-person concept proposed by the philosopher Deguchi [20,21] from an engineering viewpoint into a form that can be put into practice. The goal is to solve the ethical problems in the social application of conventional AI, which is heavily focused on intelligence, by using artificial agents as equal partners. In this section, we explain the approach to building the e-person architecture proposed in this paper.

Artificial agents that live with us and form a society with us require not only the labor, computational power, and reasoning ability that have been required of conventional artificial intelligence, but also the various social cognitive abilities that we use in our daily lives with other people. These social cognitive abilities are expected to be broad and include emotional intelligence such as empathy and compassion for others, understanding and adhering to rules and norms that may differ depending on culture and situation, and having relationships of trust and responsibility with others.

One approach to constructing e-person architecture is to consider and list these diverse cognitive abilities individually and then aggregate them (bottom-up approach). It is believed that the current solutions for AI ethics are largely based on such a bottom-up approach. However, when building an artificial agent with such a wide range of abilities in a bottom-up manner, difficulties are expected in integrating multiple cognitive abilities in a consistent manner. For example, when there is a conflict between an emotional decision based on empathy for others and a rational decision based on an understanding of laws and regulations, the difficulty arises in determining which decision is appropriate in a given situation. This problem is expected to become more complex and challenging as the number of bottom-up integrated cognitive capabilities increases. In addition, since machine learning models are black boxes and difficult to control in detail from the outside, it is expected that it will not be easy to connect machine learning models with different capabilities when using machine learning models to implement e-person. We also believe that the essence of AI ethics lies in the conflicts between these multiple decision factors. Therefore, this paper adopts a top-down construction approach, defining the capabilities that form the basis of the sociality and ethics that e-person possess, and defining different capabilities based on these capabilities.

There are several ways to think about the cognitive abilities that form the basis of the top-down definition of the e-person architecture, however we have chosen to base it on trust. The reason we base it on trust is because we believe that trust in relationships is the foundation for people to coexist with others rather than live independently. There is also a broad discussion of the capacity for trust, including emotional trust and trust in norms, and we believe it is possible to discuss the cognitive capacities expected of e-person in an integrated way.

Because trust itself is a broad topic of discussion, the top-down approach we take requires that we define it in a way that allows us to discuss such diverse types of trust in a unified way and in a format that can be manipulated in an engineering-like manner. In this paper, we use the definition of trust in social psychology as a reference for such a definition of trust, and define "the ability to cope with uncertainty" as a fundamental capability [46]. In [46], trust, which is the foundation of social relationships, is defined as follows

An individual may be said to have trust in the occurrence of an event if he expects its occurrence and his expectation leads to behavior which he perceives to have greater negative motivational consequences if the expectation is not confirmed than positive motivational consequences if it is confirmed.

In our daily lives, we are constantly exposed to diverse uncertainties, and we are always required to take these uncertainties into account when making decisions or taking actions. In real life, it is impossible to try to reduce uncertainty to zero before making a decision or taking action, because action and decision-making are required in real time. In other words, we must accept uncertainty and accept that unpredictable events may occur when making decisions or taking actions. This definition of trust in social psychology is that trust is the ability to act and make decisions under such uncertainty.

This paper considers that emotional intelligence skills such as sympathy and empathy for others, and the activities of groups in forming and sharing rules and norms, can also be seen as the ability to act and make decisions under uncertainty. For example, we can think of empathy and sympathy as a way to understand each other's uncertain hidden states and to act and make decisions together under uncertainty about others by supporting each other. We can also think of the formation and sharing of rules as a way of reducing uncertainty by sharing and clarifying appropriate behavior for each situation in advance.

Uncertainty can be treated mathematically based on probability theory, so it is also appropriate for engineering operations. The idea of viewing behavior and decision making under uncertainty as a unifying basis for cognitive ability is proposed in cognitive science as the free energy principle. A discussion of implementation based on the free energy principle is given in Section 4.

### 3.2   Definition of e-person Architecture

Based on the ideas in the previous section, this paper defines the e-person architecture by considering handling uncertainty as a fundamental cognitive ability that e-person should possess. The definition is given from two perspectives: the person perspective, which organizes uncertainty scenes comprehensively, and the level perspective, which is based on the depth of information in each person.

In our daily lives, we are confronted with various uncertainties. In order to comprehensively list these uncertainties, we classify them into the following three types based on the perspective of the person who is aware of the uncertainty faced in daily life. The three classification scenes are shown in Fig. 1.

(1)   First-person perspective uncertainty (subjective)
  Uncertainty from one party (I) towards another party (You) with whom they are face-to-face.
(2)   Second-person perspective uncertainty (intersubjective)
A bidirectional uncertainty shared by the party (I) and the other party (You), who is the other party facing them.
(3)   Third-person uncertainty (objective)
   Uncertainty from those who are not involved (They) from those who are involved (I/You).

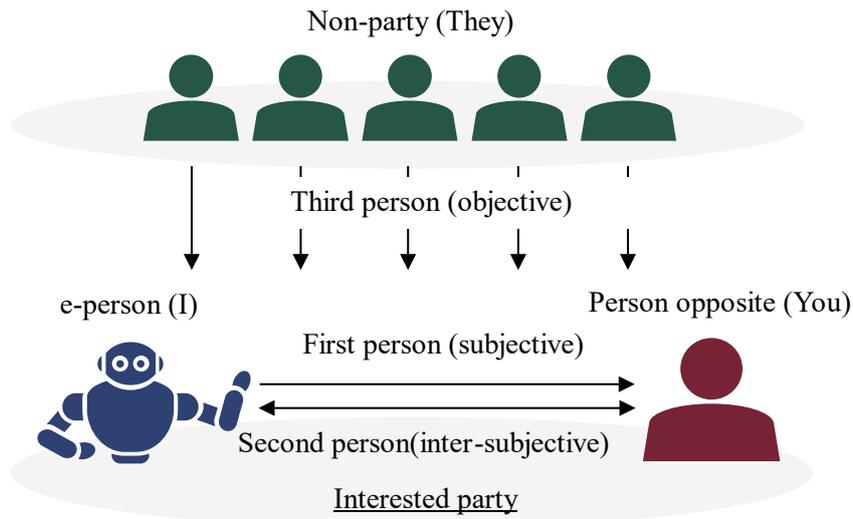

**Fig. 1.** Classification from the perspective of recognizing uncertainty

This classification is defined by the subject of the uncertainty and the direction of the uncertainty. We consider three types of uncertainty: the e-person itself (I) as the subject of uncertainty, and two or more other parties that interact directly with the subject of uncertainty: you (You), and other parties that are not the subject of uncertainty but can indirectly affect the subject of uncertainty: they (They). As shown in Fig. 1, first-person uncertainty and third-person uncertainty are defined as unidirectional uncertainty, while second person uncertainty is defined as bidirectional uncertainty.

Unidirectional uncertainty is uncertainty in a one-way inference from the source of the inference to the target of the inference. Bidirectional uncertainty, on the other hand, assumes that two or more agents are both the source and the target of the inference, and that the situation is one in which they handle the uncertainty together through communication with each other, rather than independently. These three types of uncertainty can be understood in terms of subjectivity, intersubjectivity (also called co-subjectivity or mutual subjectivity), and objectivity. First-person uncertainty is uncertainty based on inference about the environment, and can be understood as subjectivity. Similarly, objectivity is uncertainty that is given to the parties involved from the outside as common understanding or knowledge, and uncertainty that is shared by two or more agents can be understood as intersubjectivity. Note that although third person uncertainty and objectivity are uncertainties shared by multiple agents, they are not bidirectional, unlike second person uncertainty and intersubjectivity. They are given to the parties as general facts, and the parties cannot directly express their agreement or disagreement with them. Objection and agreement are always expressed to the parties, and this paper considers them as a two person scene.

Existing AI research often discusses first-person (subjective) or third-person (objective) uncertainty. For example, the world model in machine learning is a prediction model for the world based on the experience of a particular artificial agent [47]. It is assumed that the situation can be handled by simply improving the accuracy of the prediction model for the world (i.e. environment). On the other hand, for example, large language models are assumed to learn an understanding that is close to objective from the large amount of experience of third parties that is not their own experience. Large language models are assumed to acquire general knowledge that can be widely and generally applied through the vast amounts of data from different people in different scenes. Since communication between parties with different subjectivities also affects each other's subjectivities, it is also important to focus on second person scenes in the formation of subjectivities. The e-person architecture we propose focuses on second person intersubjective uncertainty and relates it to first and third person uncertainty.

One of the features of the proposed architecture is its focus on bidirectional second person uncertainty. Most discussions in conventional AI ethics have focused on questions such as "Does AI understand and follow human rules (norms or laws)?" (Ethics of AI) or "What are the negative effects of AI on human ethics?" (Ethics by AI). In contrast, our architecture considers ethics as a concept that is not a unidirectional concept from humans to AI, but rather a concept that is co-constructed through human-AI interaction (ethics with AI). With this approach, the artificial agent proposed in this paper aims to create a new relationship with artificial agents as partner, replacing the traditional use of AI as a tool. In this way, AI is not a slave that is forced by humans to deal with ethical concerns and risk measures unilaterally, but can be said to be a co-adventurer that works with humans to deal with ethical concerns and risk measures.

Next, we will organize the e-person architecture from the perspective of uncertainty based on person perspective and classify it into the following three stages. These stages are defined based on the depth of information.

(1) Uncertainty of observable information (e.g. behavior, speech, etc.)
(2) Uncertainty of shallow latent information (e.g. emotions, attention, goal, etc.)
(3) Uncertainty of deep latent information (e.g. purpose, values, etc.)

In other words, it is a classification based on whether the object of uncertainty is observable (observable information) or not (hidden information) and, in the case of unobservable hidden information, the depth of the information. Observable information is information that can be obtained directly by the observer. This includes, for example, the behavior or speech of others, which can be obtained by sight or by hearing. Conversely, unobservable hidden information includes such as other people's emotions, goals, and values. It is not possible to directly observe emotions, goals, or values. Even if another person clearly states that "I am happy." it is impossible to know whether this is true or not. For this reason, such as emotion, goals, values, etc. are hidden information. The levels are defined based on the idea that it is generally expected that hidden information is more difficult to infer than observable information, and that the deeper the level, the more difficult it is to infer. By defining these levels, we aim to support the realization of a step-by-step development of the big picture of ethics by setting development goals.

In accordance with the above two classification perspectives of the target uncertainty, the person perspective and the depth of information, the following Table 1 shows the definition of the uncertainty that the e-person architecture should address. The depth of information is considered to correspond to the difficulty of realizing e-person and is labeled as Level (Lv.) in Table 1.

In Table 1, uncertainty is classified into a total of nine types, consisting of three types based on person and three types based on information type. Each uncertainty is expressed by $u(\cdot)$. The uncertainty $u(\cdot)$ can be applied to various functions, such as the standard deviation ($\sigma$) of a probability distribution, as well as the free energy, which will be discussed in the next section. The observation information is represented as $o$, the action as $a$, and the hidden state as $s$ ($l$: $low$, $h$: $high$). The party itself is represented as $I$, the person they are talking to as $You$, and a third party as $They$. For example, Level 1 first-person uncertainty is expressed as $u_I(o_I)$ and expresses the uncertainty of the party (I) about the observational information. Similarly, Level 1 second-person uncertainty is expressed as $u_I(o_I, o_{you})$ and $u_{you}(o_{you}, o_I)$, and expresses the uncertainty of each of the two parties (I/You) regarding their own observations.

The e-person architecture is expected to handle each of the uncertainties shown in the table. For example, dealing with second-person uncertainty means reducing the two uncertainties $u_I(o_I, o_{you})$ and $u_{you}(o_{you}, o_I)$ sequentially through communication and collaboration. More rigorous probability expressions of uncertainty at the implementation level and specific implementation proposals for dealing with it are described in the framework in Section 4. As shown in Table 1, both first-person and second-person uncertainties include uncertainty about $o_I$, and they influence each other. In other words, two parties manipulating second-person uncertainty can affect each other's first-person uncertainty through communication. It is important to note that the uncertainties shown in the table can affect each other in this way. Moreover, the effect is to affect each other through communication between second person artificial agents.

**Table 1.** Definition of e-person architecture.

| | First person (subjective) | Second person (inter subjective) | Third person (objective) |
|---|---|---|---|
| Level 1: Observable (action, speech) | Uncertainty of my observation $u_I(o_I)$ | Uncertainty of shared my and your observation $u_I(o_I, o_{You}), u_{You}(o_{You}, o_I)$ | Uncertainty of shared my, your and their observation $u_I(o_I, o_{You}, o_{They})$, $u_{You}(o_{You}, o_I, o_{They})$, $u_{They}(o_{They})$ |
| Level 2: Hidden(low) (emotion, attention, goal) | Uncertainty of my hidden information $u_I(s_I^l)$ | Uncertainty of shared my and your hidden information $u_I(s_I^l, s_{You}^l), u_{You}(s_{You}^l, s_I^l)$ | Uncertainty of shared my, your and their hidden information $u_I(s_I^l, s_{You}^l, s_{They}^l)$, $u_{You}(s_{You}^l, s_I^l, s_{They}^l)$, $u_{They}(s_{They}^l)$ |
| Level 3: Hidden(high) (value, purpose) | Uncertainty of my hidden information those are higher than Level 2 $u_I(s_I^h)$ | Uncertainty of shared my and your hidden information those are higher than Level 2 $u_I(s_I^h, s_{You}^h), u_{You}(s_{You}^h, s_I^h)$ | Uncertainty of shared my, your and their hidden information those are higher than Level 2 $u_I(s_I^h, s_{You}^h, s_{They}^h)$, $u_{You}(s_{You}^h, s_I^h, s_{They}^h)$, $u_{They}(s_{They}^h)$ |

## 4 e-person Framework

### 4.1 Development Policy of the Framework

We aim to make e-persons coexist with humans. As a development strategy for e-persons coexisting with humans, referring to discussions on human-coexisting robots, there are two main candidates: the environmental structuring strategy and the humanoid strategy (so-called humanoid robots.) The environmental structuring strategy involves modifying the human living environment to match the robot's system configuration. For example, attaching two-dimensional codes to objects or environments to recognize object positions using cameras, eliminating all steps within facilities for wheeled movement, or using machine language for system input. On the other hand, the humanoid strategy involves designing the robot's system configuration to match the human living environment. For example, designing hardware to match the average height and limb structure of human adults, designing software to match human gestures, and using natural language for system input. There is no superiority or inferiority between the strategies; the difference lies in whether humans adapt to e-persons or e-persons adapt to humans as the first step towards coexistence. Regardless of the strategy taken, it is expected that as coexistence progresses, both humans and e-persons will adapt to each other. Generally, it is considered that the systems adapt to humans faster than humans adapt to the systems. For example, early computers were only used by a limited number of engineers, and before they spread to non-engineers, computers evolved to be usable by non-engineers. Considering that systems adapt faster than humans, we adopt the humanoid strategy as a development strategy for e-persons to penetrate human society more quickly.

The humanoid strategy for e-persons involves designing the system configuration of e-persons by emulating human brain functions. As mentioned in Section 2, conventional models emulating individual functions of the human brain have been designed to address individual tasks. However, considering that the human brain is a massive structure connected by generalized neurons rather than specialized structures for individual functions, it is necessary to consider not only individual functions but also the coordination and integration between functions. The system configuration of e-persons needs to be designed using a unified method that emulates the integrative nature of the human brain.

A hypothesis explaining the human brain in a unified manner is the Free Energy Principle (FEP) [48–52]. FEP is said to be a principle followed by living organisms, including humans, and unifies the perceptual and motor processes, which have been treated separately, as the minimization of variational free energy. According to FEP, variational free energy is expressed as the sum of the KL divergence between the posterior distribution $p(o|s)$ concerning the hidden state $s$ under observation $o$ and the recognition distribution $q(s)$ approximating it, and the Shannon surprise of the sensory signal:

$$F(q,p;o) = D_{KL}\big(q(s)||p(s|o)\big) + (-\log p(o)) \tag{1}$$

The perceptual process is the process of inferring the hidden state $s$ of the external environment from the observation $o$, corresponding to the minimization of the first term in equation (1). On the other hand, the motor process is the process of inferring the policy $\pi$ of actions that change the hidden state $s$ of the external environment through changes in observation $o$, corresponding to the minimization of the second term in equation (1). Therefore, the perceptual and motor processes are treated uniformly as the minimization of free energy.

FEP allows handling uncertainty by treating perceptual and motor processes uniformly as inference. Variational free energy deals with inference based on past and present observations. As mentioned in Section 3, to handle deep level uncertainty, it is necessary to infer including future observations. In FEP, the expected free energy obtained by extending variational free energy to future observations is:

$$G(\pi) = -\mathbb{E}_{Q(\tilde{s},\tilde{o}|\pi)}\big[D_{KL}[Q(\tilde{s}|\tilde{o},\pi)||Q(\tilde{s}|\pi)]\big] - \mathbb{E}_{Q(\tilde{o}|\pi)}(\ln P(\tilde{o}|C)) \tag{2}$$

Here, $\pi$ is the policy, and inferring the policy $\pi$ that minimizes the expected free energy $G$ is action planning. In the equation (2), the first term is called epistemic value, and the second term is called pragmatic value. Epistemic value is the value of seeking new information, and pragmatic value is the value of seeking favorable observations. Minimizing the expected free energy $G$ balances these values, resolving the exploration-exploitation dilemma. In other words, by increasing epistemic value, exploratory actions are taken, while by increasing pragmatic value, exploitative actions are taken, responding to environmental uncertainty.

FEP has mainly dealt with first-person agents. For example, in the behavior of a single rat in a T-maze, the position of the rat and the valence of stimuli are defined as observations, and the generative model is described. Also, in the resonant behavior of two birds, the pitch of the sounds each bird hears is defined as observations, and the generative model is described. In this way, even when dealing with multiple agents, other agents are treated without distinction from the environment, and the first-person perspective of the self-agent is modeled. To handle the uncertainty of self-agents and other agents as mentioned in Section 3, it is necessary to extend FEP to a framework that also handles second-person and third-person perspectives.

### 4.2 Framework

We explain our proposed framework using graphical models. We present models for the first person, second person, and third person.

The first person model is a hierarchical structure based on the conventional FEP. Fig. 2 shows the graphical model for the first person. The set of generative models for levels 1 and 2 in the first person is the same. The initial hidden state $s_1^l$ is inferred under the prior distribution $D$. Based on the policy $\pi$ inferred under the expected free energy $G$, the hidden state $s_\tau^l$ after transition through the transition model $B$ and the observation $o_\tau^l$ through the observation model $A$ are sequentially output. At this time, level 1 deals with the uncertainty $u(o_t^l)$ regarding the observation $o_t^l$, and level 2 deals with the uncertainty $u(s_t^l)$ regarding the hidden state $s_t^l$. By formulating the observation as the conditional probability of the hidden state represented by the observation model $A$, the uncertainty $u(o_t^l)$ regarding the observation $o_t^l$ is handled. Furthermore, by formulating the hidden state as the conditional probability of the previous hidden state and policy represented by the transition model $B$ and expected free energy $G$, the uncertainty $u(s_t^l)$ regarding the hidden state $s_t^l$ is handled. Choosing actions to minimize the expected free energy $G$ is equivalent to minimizing these uncertainties. The set of generative models for level 3 in the first person adds a hierarchy above the set of generative models for levels 1 and 2. The generative models of the higher and lower hierarchies are connected by inferring the lower hierarchy's expected free energy $G$ and prior distribution $D$ based on the higher hierarchy's observation $o_t^h$. At this time, level 3 deals with the uncertainty $u(s_t^h)$ regarding the hidden state $s_t^h$. By formulating the hidden state as the conditional probability of the previous hidden state and policy represented by the transition model $B$ and expected free energy $G$, the uncertainty $u(s_t^h)$ regarding the hidden state $u(s_t^h)$ is handled. Similar to levels 1 and 2, choosing actions to minimize the expected free energy $G$ is equivalent to minimizing this uncertainty.

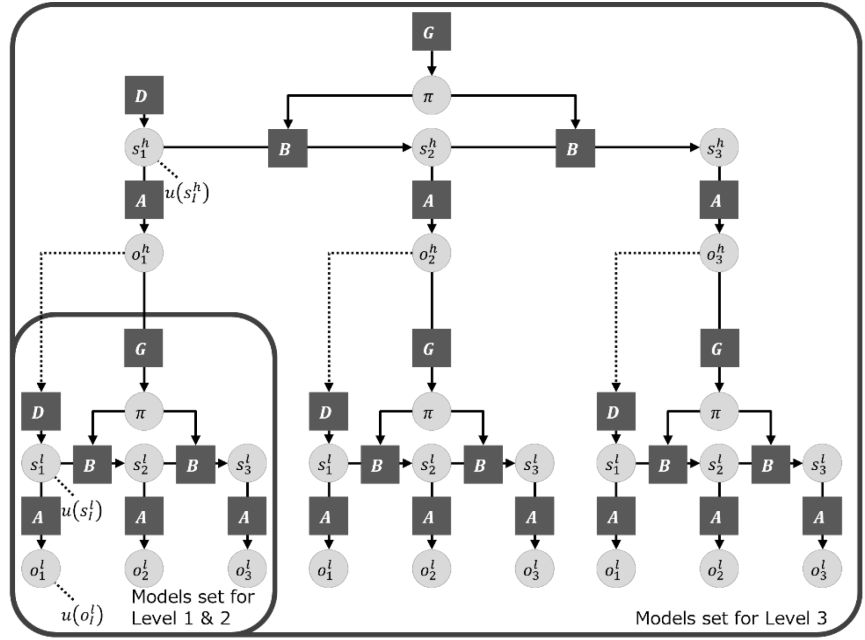

**Fig. 2.** Graphical model of first person

The second person model connects the models of two agents in the conventional FEP. Fig. 3 shows the graphical model for the second person. The set of generative models for levels 1 and 2 in the second person is the same. Unlike the first person model, the set of generative models for "I" and "You" are connected bidirectionally. Specifically, the observation model $A$ for "I" receives the hidden state $s_I$ of "I" and the hidden state $s_{You}$ of "You." Similarly, the observation model $A$ for "You" receives the hidden state $s_{You}$ of "You" and the hidden state $s_I$ of "I." In practice, the hidden states of "I" and "You" cannot be directly input to each other. Therefore, for example, the observation model $A$ for "I" receives the hidden state $s_{You}$ of "You" inferred based on the observation $o_I$ obtained by sensors of "I." A few examples will be illustrated in Section 4.3. At this time, the uncertainties handled at each level are for both "I" and "You." Level 1 deals with the uncertainty $(o_I^l, o_{You}^l)$ regarding the observation, and level 2 deals with the uncertainty $u(s_I^l, s_{You}^l)$ regarding the hidden state. Similar to the first person model, these uncertainties are handled by formulating the observation and hidden state as conditional probabilities represented by the observation model $A$, transition model $B$, and expected free energy $G$. However, unlike the first person model, the marginal distribution includes not only the hidden state $s_I$ of "I" but also the hidden state $s_{You}$ of "You." The set of generative models for level 3 in the second person adds a hierarchy above the set of generative models for levels 1 and 2. Similar to levels 1 and 2, the set of generative models for "I" and "You" are connected. Specifically, the observation model $A$ for the higher hierarchy of "I" receives the hidden state $s_I^h$ of "I" and the hidden state $s_{You}^h$ of "You." Furthermore, the calculation of the expected free energy $G$ for the lower hierarchy of "I" uses the observation $o_I^h$ of the higher hierarchy of "I" and the observation $o_{You}^h$ of the higher hierarchy of "You." At this time, level 3 deals with the uncertainty $u(s_I^h, s_{You}^h)$ regarding the hidden state. Similar to the first person model, the hidden state is formulated as the conditional probability represented by the transition model $B$ and expected free energy $G$. However, unlike the first person model, the marginal distribution includes not only the hidden state $s_I$ of "I" but also the hidden state $s_{You}$ of "You."

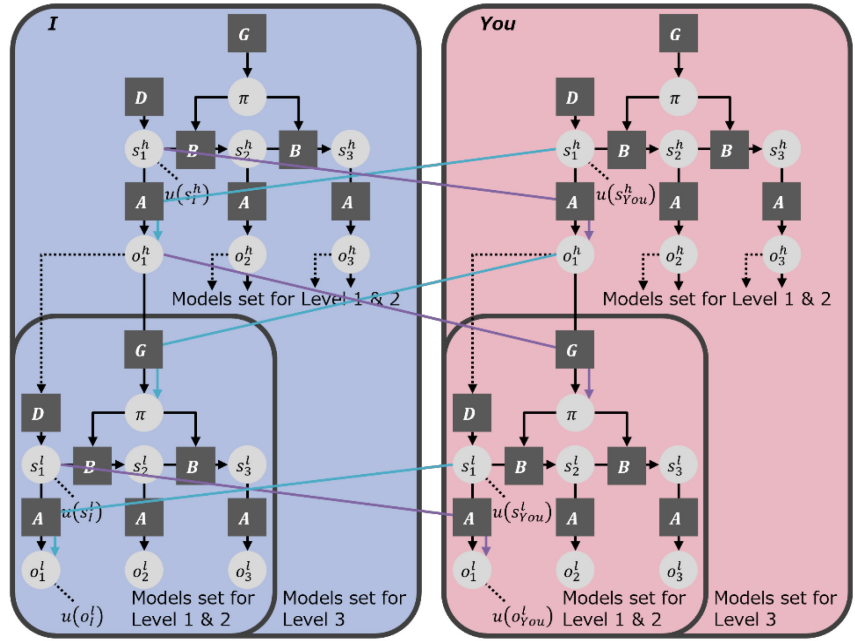

**Fig. 3.** Graphical model of second person

The third person model connects the models above the second person model, but the direction of connection is different. Fig. 4 shows the graphical model for the third person. The set of generative models for levels 1 and 2 in the third person is the same. The set of generative models for "I" and "You" are connected to the set of generative models for "They," but the connection is unidirectional, unlike the second person model. Specifically, the observation models for "I" and "You" receive the hidden states of "I" and "You" and the hidden state of "They," but the observation model for "They" only receives the hidden state of "They." At this time, the uncertainties handled at each level are for "I," "You," and "They." Level 1 deals with the uncertainty $u(o_I^l, o_{You}^l, o_{They}^l)$ regarding the observation, and level 2 deals with the uncertainty $u(s_I^l, s_{You}^l, s_{They}^l)$ regarding the hidden state. Similar to the first and second person models, these uncertainties are handled by formulating the observation and hidden state as conditional probabilities represented by the observation model $A$, transition model $B$, and expected free energy $G$. However, the marginal distribution includes not only the hidden states of "I" and "You" but also the hidden state of "They." The set of generative models for level 3 in the third person adds a hierarchy above the set of generative models for levels 1 and 2. Similar to levels 1 and 2, the set of generative models for "I" and "You" are connected unidirectionally to the set of generative models for "They." Specifically, the observation model $A$ for the higher hierarchy of "I" and "You" receives the hidden states $s_I^h$ and $s_{You}^h$ of "I" and "You" and the hidden state $s_{They}^h$ of "They." Furthermore, the calculation of the expected free energy $G$ for the lower hierarchy of "I" and "You" uses the observation $o_I^h$ and $o_{You}^h$ of the higher hierarchy of "I" and "You" and the observation $s_{They}^h$ of the higher hierarchy of "They." On the other hand, the observation model $A$ for "They" only receives the hidden state $s_{They}^h$, and the calculation of the expected free energy $G$ for "They" only uses the observation $o_{They}^h$. At this time, level 3 deals with the uncertainty $u(s_I^h, s_{You}^h, s_{They}^h)$ regarding the hidden state. Similar to the first and second person models, the hidden state is formulated as the conditional probability represented by the transition model $B$ and expected free energy $G$. However, the marginal distribution includes not only the hidden states of "I" and "You" but also the hidden state of "They."

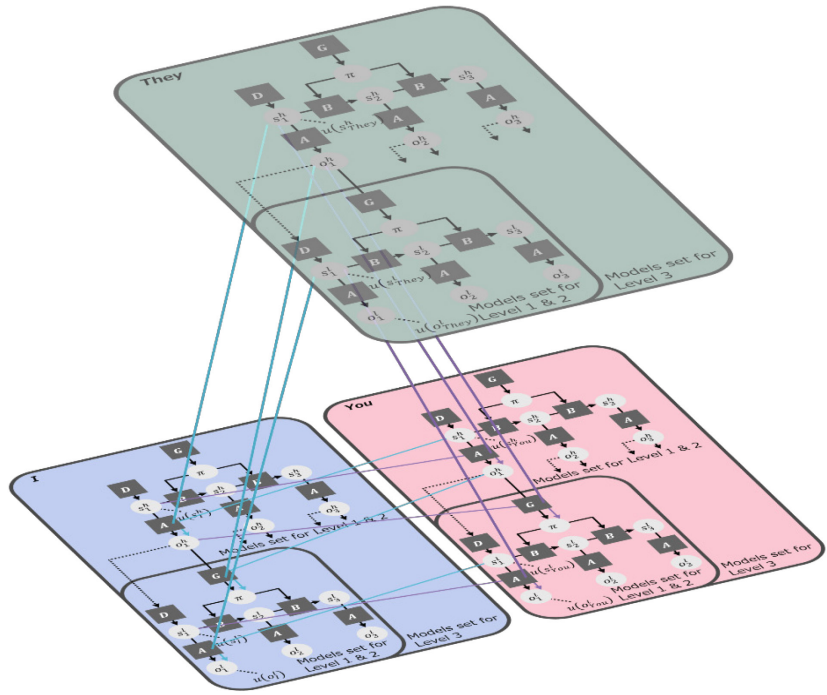

**Fig. 4.** Graphical model of third person

As shown above, the proposed framework supports multi-agent systems with dependent relationships, unlike the conventional FEP, which was mainly limited to single-agent or independent multi-agent systems. The dependent relationships between agents are formulated as mutual connections of each agent's probability distributions and generative models. Additionally, the connection between "I" and "You" is bidirectional, while the connection with "They," who are positioned above the agents, is unidirectional, encompassing different types of dependent relationships. Furthermore, the ease of observation associated with dependent relationships is expressed through the hierarchy from levels 1 to 3. By seamlessly connecting directly observable low-level observations to high-level hidden states that are not directly observable, it is possible to implement a system that handles something that is not directly observable, such as values.

## 4.3  Application Examples of the Framework

To handle first-person, level 1 uncertainty, the e-person framework was applied to partial observation character recognition [53] and object manipulation with dynamic properties [54]. Partial observation character recognition involves identifying characters by moving the observation area in situations where the entire character cannot be observed at once. The uncertainty in this problem arises from not knowing what character it is due to the unknown image outside the observable area. By applying the e-person framework, the system sequentially inferred characters and determined the observation area to move to. On the other hand, object manipulation with dynamic properties involves storing objects in a box in situations where it is unknown how the object will deform until grasped by a robot. The uncertainty in this problem arises from not knowing the trajectory for storing the object in the box due to unknown object behavior. By applying the e-person framework, the robot successfully stored objects in a box by executing trajectories with margins to avoid collisions with surrounding obstacles based on the inference accuracy of object behavior. In both cases, uncertainty was addressed by accumulating better actions from a first-person perspective.

To handle first-person, level 2 uncertainty, the e-person framework was applied to dynamic selection of robot joints according to objectives [55]. Dynamic selection of robot joints according to objectives involves appropriately selecting joints necessary for a robot to perform a target task. The uncertainty in this problem arises from not knowing which joints to move due to an underdetermined problem where the degrees of freedom for motion required to perform the target task are smaller than the number of joints. By applying the e-person framework, we transformed it into a problem of which robot joints to focus on and performed the target task by moving only the minimum necessary joints. Uncertainty was addressed by focusing on embodiment from a first-person perspective.

To handle second-person, level 1 uncertainty, the e-person framework was applied to autonomous movement of robots at scramble intersections[56,57] and collaborative transport by robots with different embodiments[58]. Autonomous movement of robots at scramble intersections involves four robots simultaneously approaching an intersection from different directions and autonomously moving in such situations. The uncertainty in this problem arises from not knowing which trajectory to take due to unknown trajectories of three other robots besides one's own robot. By applying the e-person framework, the trajectories for one's own robot are generated based on inference of one's own robot's trajectory seen from other robots. On the other hand, collaborative transport by robots with different embodiments involves two robots with different reachable areas cooperating to transport objects in situations where a single robot cannot transport objects to target positions. The uncertainty in this problem arises from not knowing when one's own robot should act. By applying the e-person framework, we determined one's own robot's action timing without pre-teaching timing by inferring expected actions of other robots along with one's own robot's actions. In both cases, uncertainty was addressed by deciding actions from a second-person perspective regarding relationships with others. Thus, even when applying the same second-person e-person framework problem, implementation methods differ depending on how boundaries between "I" and "You" are determined. The e-person framework has implementation flexibility and potential applicability to more problems.

To handle second-person, level 2 uncertainty, the e-person framework was applied to public goods games with penalties[59]. Public goods games with penalties involve two players deciding whether or not to contribute their resources and whether or not to penalize their opponent player. The uncertainty in this problem arises from not knowing what actions one's opponent player will take due to unknown opponent behavior. By applying the e-person framework, the system balanced expected free energy of others and oneself as social emotional value and decided one's own actions so that this value becomes neutral. Uncertainty was addressed by inferring emotions while deciding actions from a second-person perspective regarding relationships with others.

We have previously applied the e-person framework to observable and unobservable and shallow uncertainties for first-person and second-person perspectives. We have not yet addressed unobservable and deep uncertainties for first-person and second-person perspectives regarding goals and values. To apply the e-person framework to these uncertainties, it is necessary to consider mechanisms for providing goals and values. One direction is sharing goals and values from others besides oneself (such as past selves or current partners). This is an extrinsic perspective on goals and values. Additionally, we have not yet addressed third-person perspective either. To apply the e-person framework to third-person perspective, it is necessary to consider mechanisms for acquiring third-person perspective. One direction is treating agents based on large language models (LLMs), which are currently attracting attention, as third persons.

## 5  Conclusion

We proposed the e-person architecture with the aim of constructing a unified foundation to address AI ethics issues and its incremental development. The e-person architecture is a unified foundation that reduces uncertainty through collaborative cognition and action with others, i.e., "co-adventure." By classifying and defining uncertainty along two axes: (1) the perspectives of first person, second person, and third person, and (2) the uncertainty based on the latent nature of information at three levels, we support the incremental development of the foundation. Additionally, to implement the e-person architecture, we proposed the e-person framework based on the free energy principle, which considers the reduction of uncertainty as a unified principle of brain function.

We presented our application examples of the proposed framework and future challenges. Our contributions are as follows:

1. We presented the e-person architecture that supports the incremental development of e-person proposed by Deguchi et al., using uncertainty as a unified metric that can be quantitatively handled.
2. We demonstrated the e-person framework to implement the e-person architecture based on the free energy principle.
3. We applied the e-person framework to a wide range of tasks, from character recognition to autonomous robot navigation and collaborative transport by multiple robots, showing that the e-person architecture serves as a unified foundation and contributes to incremental development.

**Acknowledgments.** This research was supported by JST RISTEX Grant Number JPMJRS22J5 and JSPS Topic-Setting Program to Advance Cutting-Edge Humanities and Social Sciences Research Grant Number JPJS00122679495.

**Disclosure of Interests.** The authors have no competing interests to declare that are relevant to the content of this article.

## References

1. Future of life institute: Asilomar AI principle. https://futureoflife.org/open-letter/ai-principles/ (last access 2025/02/26)
2. Bostrom, Nick, and Eliezer Yudkowsky. "The ethics of artificial intelligence." Artificial intelligence safety and security. Chapman and Hall/CRC, 2018. 57-69.
3. Floridi, Luciano, et al. "AI4People—an ethical framework for a good AI society: opportunities, risks, principles, and recommendations." Minds and machines 28 (2018): 689-707.
4. Mittelstadt, Brent. "Principles alone cannot guarantee ethical AI." Nature machine intelligence 1.11 (2019): 501-507.
5. Jobin, Anna, Marcello Ienca, and Effy Vayena. "The global landscape of AI ethics guidelines." Nature machine intelligence 1.9 (2019): 389-399.
6. Hagendorff, Thilo. "The ethics of AI ethics: An evaluation of guidelines." Minds and machines 30.1 (2020): 99-120.
7. Floridi, Luciano, and Josh Cowls. "A unified framework of five principles for AI in society." Machine learning and the city: Applications in architecture and urban design (2022): 535-545.
8. Kurzweil, Ray. "The singularity is near." Ethics and emerging technologies. London: Palgrave Macmillan UK, 2005. 393-406.
9. Bostrom, Nick. "Superintelligence: Paths, dangers, strategies." Oxford University Press (2014).
10. European Union: Regulation (EU) 2024/1689 of the European Parliament and of the Council of 13 June 2024 laying down harmonised rules on artificial intelligence and amending Regulations (EC) No 300/2008, (EU) No 167/2013, (EU) No 168/2013, (EU) 2018/858, (EU)

2018/1139 and (EU) 2019/2144 and Directives 2014/90/EU, (EU) 2016/797 and (EU) 2020/1828 (Artificial Intelligence Act)
11. Xu, Feiyu, et al. "Explainable AI: A brief survey on history, research areas, approaches and challenges." Natural language processing and Chinese computing: 8th cCF international conference, NLPCC 2019, dunhuang, China, October 9–14, 2019, proceedings, part II 8. Springer International Publishing, 2019.
12. Angelov, Plamen P., et al. "Explainable artificial intelligence: an analytical review." Wiley Interdisciplinary Reviews: Data Mining and Knowledge Discovery 11.5 (2021): e1424.
13. Bellamy, Rachel KE, et al. "AI Fairness 360: An extensible toolkit for detecting and mitigating algorithmic bias." IBM Journal of Research and Development 63.4/5 (2019): 4-1.
14. Mehrabi, Ninareh, et al. "A survey on bias and fairness in machine learning." ACM computing surveys (CSUR) 54.6 (2021): 1-35.
15. IBM, "2023 ESG REPORT IBM impact report", 2023.
16. Corbett-Davies, Sam, et al. "The measure and mismeasure of fairness." Journal of Machine Learning Research 24.312 (2023): 1-117.
17. Pessach, Dana, and Erez Shmueli. "A review on fairness in machine learning." ACM Computing Surveys (CSUR) 55.3 (2022): 1-44.
18. SAE international, Taxonomy and Definitions for Terms Related to Driving Automation Systems for On-Road Motor Vehicles J3016_202104
19. Morris, M.R., Sohl-Dickstein, J., Fiedel, N., Warkentin, T., Dafoe, A., Faust, A., Farabet, C. & Legg, S.. (2024). Position: Levels of AGI for Operationalizing Progress on the Path to AGI. Proceedings of the 41st International Conference on Machine Learning, in Proceedings of Machine Learning Research 235:36308-36321
20. 出口康夫, "AI親友論", 徳間書店, 2023. (in Japanese)
21. Yasuo Deguchi, "Creating a "We" Society in which Humans ("I") and AIs with Personhood ("e-people") Coexist", in Hitachi Review special article, report Sixth Symposium of Kyoto University and Hitachi Kyoto University Laboratory (2023).
22. Floridi, Luciano, and Jeff W. Sanders. "On the morality of artificial agents." Minds and machines 14 (2004): 349-379.
23. Moor, James. "The Dartmouth College artificial intelligence conference: The next fifty years." AI magazine 27.4 (2006): 87-87.
24. Nick, Bostrom. "Superintelligence: Paths, dangers, strategies." 2014,
25. Russell, Stuart. Human compatible: AI and the problem of control. Penguin Uk, 2019.
26. Cappelen, Herman, and Josh Dever. Making AI intelligible: Philosophical foundations. Oxford University Press, 2021.
27. Gao, Di Kevin, et al. "AI ethics: a bibliometric analysis, critical issues, and key gaps." International Journal of Business Analytics (IJBAN) 11.1 (2024): 1-19.
28. Velthoven, Max, and Eric Marcus. "Problems in AI, their roots in philosophy, and implications for science and society." arXiv preprint arXiv:2407.15671 (2024).
29. Benjamins, V. Richard, and Idoia Salazar García. "Towards a framework for understanding societal and ethical implications of Artificial Intelligence." Vulnerabilidad y cultura digital: riesgos y oportunidades de la sociedad hiperconectada (2019): 89-100.
30. Deranty, Jean-Philippe, and Thomas Corbin. "Artificial intelligence and work: a critical review of recent research from the social sciences." Ai & Society 39.2 (2024): 675-691.
31. Capraro, Valerio, et al. "The impact of generative artificial intelligence on socioeconomic inequalities and policy making." PNAS nexus 3.6 (2024): pgae191.
32. Pedreschi, Dino, et al. "Human-AI coevolution." Artificial Intelligence (2024): 104244.

33. Mahmud, Bahar, Guan Hong, and Bernard Fong. "A study of human–ai symbiosis for creative work: Recent developments and future directions in deep learning." ACM Transactions on Multimedia Computing, Communications and Applications 20.2 (2023): 1-21.
34. Xu, Ruoxi, et al. "AI for social science and social science of AI: A survey." Information Processing & Management 61.3 (2024): 103665.
35. Solís, M. W. M. V., et al. "The impact of artificial intelligence on higher education: A sociological perspective." Journal of Namibian Studies: History Politics Culture 33 (2023): 3284-3290.
36. Breazeal, Cynthia. "Toward sociable robots." Robotics and autonomous systems 42.3-4 (2003): 167-175.
37. Dzindolet, Mary T., et al. "The role of trust in automation reliance." International journal of human-computer studies 58.6 (2003): 697-718.
38. Lee, John D., and Katrina A. See. "Trust in automation: Designing for appropriate reliance." Human factors 46.1 (2004): 50-80.
39. Goodrich, Michael A., and Alan C. Schultz. "Human–robot interaction: a survey." Foundations and trends® in human–computer interaction 1.3 (2008): 203-275.
40. Gray, Kurt, and Daniel M. Wegner. "Feeling robots and human zombies: Mind perception and the uncanny valley." Cognition 125.1 (2012): 125-130.
41. Amodei, Dario, et al. "Concrete problems in AI safety." arXiv preprint arXiv:1606.06565 (2016).
42. Amershi, Saleema, et al. "Guidelines for human-AI interaction." Proceedings of the 2019 chi conference on human factors in computing systems. 2019.
43. Prakash, Nikhil, and Kory W. Mathewson. "Conceptualization and Framework of Hybrid Intelligence Systems." arXiv preprint arXiv:2012.06161 (2020).
44. Fabri, Lukas, et al. "Disentangling human-AI hybrids: conceptualizing the interworking of humans and AI-enabled systems." Business & information systems engineering 65.6 (2023): 623-641.
45. Holter, Steffen, and Mennatallah El-Assady. "Deconstructing Human-AI Collaboration: Agency, Interaction, and Adaptation." Computer Graphics forum. Vol. 43. No. 3. 2024.
46. Deutsch, Morton. "Trust and suspicion." Journal of conflict resolution 2.4 (1958): 265-279.
47. Ha, David, and Jürgen Schmidhuber. "Recurrent world models facilitate policy evolution." Advances in neural information processing systems 31 (2018).
48. Friston, Karl, James Kilner, and Lee Harrison. "A free energy principle for the brain." Journal of physiology-Paris 100.1-3 (2006): 70-87.
49. Friston, Karl. "The free-energy principle: a unified brain theory?." Nature reviews neuroscience 11.2 (2010): 127-138.
50. McGregor, Simon, Manuel Baltieri, and Christopher L. Buckley. "A minimal active inference agent." arXiv preprint arXiv:1503.04187 (2015).
51. Friston, Karl, et al. "Active inference: a process theory." Neural computation 29.1 (2017): 1-49.
52. Parr, Thomas, Giovanni Pezzulo, and Karl J. Friston. Active inference: the free energy principle in mind, brain, and behavior. MIT Press, 2022.
53. K. Esaki, T. Matsumura, K. Ito, and H. Mizuno, "Sensorimotor visual perception on embodied system using free energy principle," Commun. Comput. Inf. Sci., vol. 1524, pp. 865–877, 2021.
54. K. Esaki, T. Matsumura, C. Yoshimura, and H. Mizuno, "Extended-self recognition for autonomous agent based on controllability and predictability," 2022 IEEE Symposium Series on Computational Intelligence (SSCI), pp. 1036–1043, 2022.

55. K. Esaki, T. Matsumura, S. Minusa, Y. Shao, C. Yoshimura, H. Mizuno, "Dynamical Perception-Action Loop Formation with Developmental Embodiment for Hierarchical Active Inference," Commun. Comput. Inf. Sci., vol. 1915, pp. 14–28, 2023.
56. T. Matsumura, K. Esaki, and H. Mizuno, "Empathic active inference: active inference with empathy mechanism for socially behaved artificial agent," Proc. ALIFE 2022: The 2022 Conference on Artificial Life, pp.18.
57. T. Matsumura, K. Esaki, S. Minusa, Y. Shao, C. Yoshimura, H. Mizuno, "Social Emotional Valence for Regulating Empathy in Active Inference," Proceedings of the 2023 Artificial Life Conference, 2023.
58. K.Esaki, et al., "Artificial Minimal Self on Free Energy Principle for Autonomous Cooperative Behavior." ALIFE 2024: Proceedings of the 2024 Artificial Life Conference. MIT Press, 2024.
59. T. Matsumura, K. Esaki, S. Yang, C. Yoshimura, H. Mizuno, "Active Inference With Empathy Mechanism for Socially Behaved Artificial Agents in Diverse Situations," Artif Life 2024; 30 (2): 277–297.